\documentclass[aps,prx,reprint,groupedaddress]{revtex4-1}
\usepackage[hyperindex,breaklinks]{hyperref}
\usepackage{url}
\usepackage[dvips]{graphicx}
\usepackage{amsthm}
\usepackage{amsmath}
\usepackage{physics}
\usepackage{color}
\usepackage{here}
\usepackage{amssymb}
\usepackage{dcolumn}
\usepackage{bm}
\usepackage{subfigure}
\usepackage{bbold}

\newcommand{\com}[1]{\relax{#1}}

\allowdisplaybreaks[0]

\begin{document}
\title{Collectively enhanced high-power and high-capacity charging of quantum batteries via quantum heat engines}

\author{Kosuke Ito$^{1,2}$}\email{kosuke.ito@qc.ee.es.osaka-u.ac.jp}
\author{Gentaro Watanabe$^{2,3}$}\email{gentaro@zju.edu.cn}
\affiliation{
${}^1$Center for Quantum Information and Quantum Biology, Institute for Open and Transdisciplinary Research Initiatives, Osaka University, Osaka 560-8531, Japan\\
${}^2$Department of Physics and Zhejiang Institute of Modern Physics, Zhejiang University, Hangzhou, Zhejiang 310027, China\\
${}^3$Zhejiang Province Key Laboratory of Quantum Technology and Device, Zhejiang University, Hangzhou, Zhejiang 310027, China}

\begin{abstract}
 As a model of so-called quantum battery (QB), quantum degrees of freedom as energy storage, we study a charging protocol of a many-body QB consisting of $N$ two-level systems (TLSs) using quantum heat engines (QHEs). We focus on the collective enhancement effects in the charging performance of QBs in comparison to the individual charging. It is a challenging goal of QBs to achieve large collective enhancements in the charging power and the capacity while keeping the experimental feasibility, the stability, and the cheapness of the required control and resources. We show that our model actually exhibits these features. In fact, our protocol simultaneously achieves the asymptotically-perfect charge and almost $N$-order average power enhancement with only thermal energy resource and simple local interactions in a stable manner. The capacity is collectively enhanced due to the emergent bosonic quantum statistics caused by the symmetry of the interaction between the engine and the batteries, which results in asymptotically perfect excitation of all the TLSs. The charging speed, and hence the average power are collectively enhanced by the superradiance-like cooperative excitation in the effective negative temperature. Our results suggest that QHEs actually fit for a charger of QBs, efficiently exploiting the collective enhancements, not only converting the disordered thermal energy to the ordered energy stored in quantum degrees of freedom.
\end{abstract}

\maketitle

\section{Introduction}
 As the technology to control quantum systems is developed,
energetics in quantum scale has been extensively and intensively studied in the context of quantum thermodynamics \cite{e15062100,Pekola:2015aa,sai-qthermo2015,1751-8121-49-14-143001,1367-2630-18-1-011002,1367-2630-19-1-010201,binder2019thermodynamics,Lostaglio_2019,doi:10.1080/00107514.2019.1631555}.
Specifically, the research on quantum batteries (QBs) aims at understanding how to utilize quantum degrees of freedom as
energy storage media.
Since Alicki and Fannes \cite{PhysRevE.87.042123} introduced QB,
collective effects of $N$ identical QBs
have been one of the main focus points in the literature.
A detailed review can be found in \cite{Campaioli2018}.
In general, entangling operations on many QBs can enhance their performance in comparison with the individual charging \cite{PhysRevE.87.042123, PhysRevLett.111.240401, Binder_2015, PhysRevLett.118.150601}, though the entanglement generation itself is not always mandatory \cite{PhysRevLett.111.240401, PhysRevLett.118.150601}.
The advantage of the collective operation can be geometrically interpreted \cite{Binder_2015}.
\com{Utilizing the quantum speed limit, Campaioli {\it et al.} \cite{PhysRevLett.118.150601} established upper bounds
for the collective advantage in the charging power defined as the ratio of the power obtained by the collective operation to the one obtained by the individual operation.}
Their bounds show that the highest achievable collective advantage of the charging power scales $N$ order under reasonable constraints.
Juli\`a-Farr\'e {\it et al.} \cite{PhysRevResearch.2.023113} derived a general upper bound for the charging power which characterizes the distinct contributions from the entanglement and the speed of the evolution.
Other general bounds on the charging power have also been shown in relation with the energy fluctuation \cite{PhysRevLett.125.040601}, and for open QBs \cite{zakavati_bounds_2020}.

Besides these fundamental results, how can high-performance QBs be realized in concrete systems utilizing collective enhancements?
Various models have been examined so far to address this issue \cite{PhysRevLett.120.117702,PhysRevA.97.022106,1812.10139,PhysRevB.99.035421,PhysRevE.99.052106,Liu:2019aa,PhysRevLett.122.210601,PhysRevB.100.115142,PhysRevResearch.2.023095,doi:10.1063/1.5096772,PhysRevE.100.032107,1912.07234,PhysRevResearch.2.013095,PhysRevA.101.032115}.
Locality of the operation is one of the important issues for the experimental feasibility.
However, the upper bound for the charging power given in \cite{PhysRevLett.118.150601} tells that
the power is restricted by the interaction order and the maximum participation number (the maximum number of batteries with which one battery can interact).
Thus, it is a nontrivial problem to realize an experimentally feasible high-power charging protocol.
In fact, the models of the optimal charging protocols \cite{Binder_2015, PhysRevLett.118.150601} rely on highly global interactions.
As the first proposal of the experimentally feasible charging protocol of QBs with the large scaling of the power,
Ferraro {\it et al.} \cite{PhysRevLett.120.117702} proposed a model of collective charging protocol of QBs where each two-level QB interacts with a quantum cavity mode as the charger.
This so-called Dicke QB model achieves $\sqrt{N}$-order of the collective advantage in the charging power by 
the two-body interactions with a common quantum cavity mode, which result in
the large effective participation number.
Recently, Quach and Munro \cite{PhysRevApplied.14.024092} proposed a model of QB composed of spins charged by other charger spins under the environmental dissipation.
They numerically showed that their model achieves $N$-order instantaneous charging power with small $N$.
However, it is only the maximum instantaneous power with small $N$ that the $N$-order advantage has been shown.
In fact, an experimentally feasible model of QBs with $N$-order collective advantage of the average charging power is still open to explore.
Moreover, not only the charging power, but also the capacity is an essential figure of merit of QBs.
It is another challenging goal to store large extractable energy in a stable manner \cite{1912.07247}.
Collective effects can also result in the superextensive capacity \cite{PhysRevApplied.14.024092}.
\com{Using the dark state, Quach and Munro's model \cite{PhysRevApplied.14.024092} is also favorable as a stable open QB.
Open QBs are models of QBs where QBs are treated as open systems in contrast to closed QBs where QBs are treated as closed systems.
Open QBs are recently gathering attention to address the problem of stabilization of the charging under the environmental dissipation \cite{doi:10.1063/1.5096772,zakavati_bounds_2020,PhysRevB.99.035421,Carrega_2020,2005.12823,2005.08489,2005.12859,PhysRevResearch.2.013095}.}

On the other hand, the cost of implementing the charging is critical as well.
The ideal charging protocol of QBs is such that it achieves high power and capacity with small control effort and as cheap a resource as possible.
However, conventional proposals including the above mentioned ones require
well-prepared excited quantum states of the charger
or well-controlled driving fields as the energy resources.
\c{C}akmak's recent proposal \cite{2005.08489} focuses on the cheapness of the control and resource of the charging protocol, aside from the power and the capacity.
In that model, through the collective interaction with a single heat bath, two-level systems can store the extractable amount of the energy in the steady state, solely from the heat bath due to the maintained coherence in the collective eigenbasis.
Although the control and resource requirement is cheap in that model, its capacity and power are quite small, as the extractable energy is stored in a thermal state with the finite temperature.

\com{
As for a feasible charger with cheap resource, quantum heat engine (QHE) can be a good candidate, as it extracts the work from the heat bath.
In fact, models of the work extraction by QHEs to quantum work storage systems have been studied \cite{Skrzypczyk:2014aa,Guryanova:2016aa,Masanes:2017aa,PhysRevE.96.012128,PhysRevA.95.032132,PhysRevLett.118.050601,PhysRevE.97.012129,PhysRevLett.124.210603}.
Recently, microscopic thermal machines have been experimentally implemented \cite{Rossnagel325,Maslennikov:2019aa,PhysRevLett.122.110601}.
Especially, the latest one \cite{PhysRevLett.122.110601} has demonstrated inherent quantum effects related to the internal coherence in the engine \cite{PhysRevX.5.031044}.
However, the performance of QHEs as the charger of QBs to fully utilize the collective effects
is an open issue.
}

In this paper,
we show that QHE is actually a promising charger of QBs taking into account the above issues,
by providing a model of charging QBs using QHEs which exhibits collectively enhanced high power and capacity with cheap controls and resources.
\com{Remarkably, the capacity of our model achieves maximum-possible amount (perfect charge) in the asymptotic limit of large number $N$ of QBs as a result of the collective enhancement caused by the effective quantum statistics.
The quality of the stored energy is accordingly enhanced in the sense of its fast decreasing variance as $N$ goes large.
Moreover, the average charging power reaches almost $N$-order scaling as a result of the superradiance-like collective effect.
As for the stability and the cheapness of the control and resource, the energy is stably stored using only thermal energy resource just by establishing time-independent local interactions between QHEs and QBs.}

We begin by showing the detail of our model in Sec.~\ref{sec_model}.
Then, Secs.~\ref{sec_capacity} and \ref{sec_speed} are devoted to the analysis of its capacity and the charging power respectively.
Finally, we conclude in Sec.~\ref{conclusion}, including a possible experimental realization of our model.

\section{Quantum heat engine and quantum battery}\label{sec_model}
\subsection{Model}
\begin{figure*}
\centering
 \includegraphics[clip ,width=7.0in]{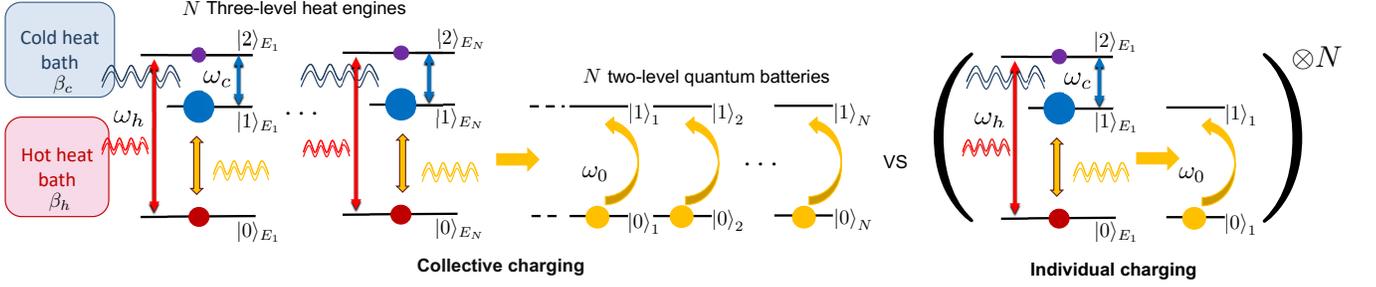}
 \caption{Schematic picture of the QHEs and the QBs.}
\label{QHEQB}
\end{figure*}
We consider a three-level heat engine similar to the three-level maser heat engine \cite{PhysRevLett.2.262},
which consists of two heat baths and a three-level system as the working substance (Fig.~\ref{QHEQB}).
The two heat baths have different respective inverse temperatures $\beta_h$ and $\beta_c$.
In this paper, we set $\hbar = k_B = 1$.
$N$ engines are used to charge $N$ quantum batteries.
The Hamiltonian of the working substance of each $j$-th engine is given by $H_{E_j} = \omega_h \ketbra{2}_{E_j} + \omega_0 \ketbra{1}_{E_j} + 0 \times \ketbra{0}_{E_j} = \omega_h \ketbra{2}_{E_j} + \omega_0 \ketbra{1}_{E_j}$, where $\{\ket{0}_{E_j}, \ket{1}_{E_j}, \ket{2}_{E_j}\}$ is the complete orthonormal basis of the Hilbert space of the three-level working substance.
We do not explicitly treat the heat baths as quantum systems.
Instead, we consider the following Lindblad-Gorini-Kossakowski-Sudarshan (LGKS) master equation \cite{Lindblad1976,doi:10.1063/1.522979} of the working substance as the resulting dynamics from the interaction with the two heat baths:
\begin{align}
 \frac{d\relax{\rho_{E_j}}}{dt}
=& -i[H_{E_j}, \relax{\rho_{E_j}}] \nonumber\\
&+\sum_{b = h, c}\Gamma_{b} \left[ \mathcal{D}[\relax{a}_{b,j}](\relax{\rho}_{E_j}) + e^{-\beta_{b} \omega_{b}} \mathcal{D}[\relax{a}_{b,j}^{\dagger}](\relax{\rho}_{E_j})\right],
\end{align}
where $\omega_c := \omega_h - \omega_0$ is the energy gap between $\ket{1}_{E_j}$ and $\ket{2}_{E_j}$, $\Gamma_b \; (b = h, c)$ is the relaxation rate with respect to each bath,
$\relax{a}_{h,j} := \ketbra{0}{2}_{E_j}$, $\relax{a}_{c,j} := \ketbra{1}{2}_{E_j}$, and
\begin{align}
 \mathcal{D}[\relax{a}](\relax{\rho}) := \relax{a}\relax{\rho}\relax{a}^{\dagger} - \frac{1}{2}(\relax{a}^{\dagger}\relax{a}\relax{\rho} + \relax{\rho}\relax{a}^{\dagger}\relax{a}).
\end{align}
In this model, the transition between the levels $\ket{0}_{E_j}$ and $\ket{2}_{E_j}$ is coupled with the hotter heat bath, and similarly, the transition between $\ket{1}_{E_j}$ and $\ket{2}_{E_j}$ is coupled with the colder bath.
The steady state $\relax{\rho}_{E_j}^{\mathrm{SS}} $of the engine under this dynamics is given as
\begin{align}
\relax{\rho}_{E_j}^{\mathrm{SS}}
 =\frac{e^{\beta_h \omega_h}\ketbra{0}_{E_j} + e^{\beta_c \omega_c}\ketbra{1}_{E_j} + \ketbra{2}_{E_j}}{e^{\beta_h \omega_h} + e^{\beta_c \omega_c} + 1}.
\end{align}
To extract the work, we impose the condition $\beta_h / \beta_c < \omega_c / \omega_h < 1$.
Indeed, this parameter regime yields the population inversion $\bra{0}\relax{\rho}_{E_j}^{\mathrm{SS}}\ket{0}_{E_j} < \bra{1}\relax{\rho}_{E_j}^{\mathrm{SS}}\ket{1}_{E_j}$ between the ground state $\ket{0}_{E_j}$ and the first excited state $\ket{1}_{E_j}$.
Then, we can extract the work to the battery through the coupling with the transition mode $\ket{0}_{E_j} \leftrightarrow \ket{1}_{E_j}$ as we explicitly do so later [Eq.~\eqref{H_col}].
This model may be realized through the respective frequency filters passing $\omega_h$, $\omega_c$, and $\omega_0$ \cite{PhysRevLett.2.262}.

As for the battery, we consider a two-level system with the resonance frequency $\omega_0$ as a unit quantum battery to be charged by the heat engine.
We consider $N$ noninteracting quantum batteries, whose Hamiltonian is $H_{B} = \sum_{j=1}^N H_{B_j} = \sum_{j=1}^{N} \omega_0 \ketbra{1}_j$, where $H_{B_j} = \omega_0 \ketbra{1}_j$ is the Hamiltonian of $j$-th battery.
The batteries are initialized in the empty state $ \ket{\psi_0} := \otimes_{j=1}^{N}\ket{0}_j$.

To investigate the collective effects, we compare the collective-charging protocol with the individual-charging protocol. Especially, the interaction between the batteries is neglected to focus on the effects purely come from the collectiveness of the charging operation itself.
In the collective-charging protocol (Fig.~\ref{QHEQB} left), the interaction
\begin{align}
 K_{j,k}^{(N)} = i\frac{g}{\sqrt{N}}(\ketbra{1}{0}_{E_j}\otimes \relax{\relax{S}}_{B_k}^{-} - \ketbra{0}{1}_{E_j} \otimes \relax{\relax{S}}_{B_k}^{+})
\end{align}
is introduced between every $i$-th engine and $j$-th battery so that all $N$ batteries are collectively charged by each heat engine, where $\relax{\relax{S}}_{B_k}^{+} = \ketbra{1}{0}_k$ and $\relax{\relax{S}}_{B_k}^{-} = \ketbra{0}{1}_k$.
Here, the coupling constant is normalized by $1/\sqrt{N}$ factor in accordance with the condition to have the well-defined thermodynamic limit for $N\rightarrow \infty$ \cite{PhysRevResearch.2.023113}.
The total Hamiltonian in this case is written as
\begin{align}
 &H_{\mathrm{tot}}^{(\mathrm{c})} \nonumber\\
=& \sum_{j=1}^{N}H_{E_j} + \sum_{k=1}^{N} H_{B_k} + \sum_{j, k=1}^{N} K_{j,k}^{(N)}\nonumber\\
=& H_{E} + H_B + \sum_{j=1}^{N} i\frac{g}{\sqrt{N}}(\ketbra{1}{0}_{E_j}\otimes \relax{\relax{S}}_{B}^{-} - \ketbra{0}{1}_{E_j} \otimes \relax{\relax{S}}_{B}^{+}),\label{H_col}
\end{align}
where $\relax{\relax{S}}_{B}^{\pm} = \sum_{j=1}^{N}\relax{\relax{S}}_{B_{j}}^{\pm}$ and $H_E = \sum_{j=1}^{N}H_{E_j}$.
\com{In experiments, we can also take the convention without the normalization factor $1/\sqrt{N}$ up to some finite $N$ which is possible to put together while keeping a priori fixed coupling strength.}
Even if we use only single engine to charge $N$ batteries in the convention without the normalization factor $1/\sqrt{N}$, the main results below do not change.
Note that this charging protocol only involves local two-body interactions, which is important for the experimental feasibility.
In addition, the energy resource in our model is purely heat from the reservoir, requiring no prefilled charger quantum state
or time-dependent control of the Hamiltonian
like in conventional models.
This protocol is cheap in this sense.
We compare the scaling of the performance of this collective-charging protocol with the individual-charging protocol (Fig.~\ref{QHEQB} right).
In the individual-charging protocol, the interaction $K_{j,j}^{(1)}$ is only imposed between the engine and battery with the same index $j$.
Hence, each single battery is charged by a single engine.
Here, the coupling constant is not normalized by $1/{\sqrt{N}}$ in this case reflecting the fact that the pairs of an engine and a battery are independent from each other.
The total Hamiltonian in this case reads
\begin{align}
 &H_{\mathrm{tot}}^{(\mathrm{i})} \nonumber\\
=& \sum_{j=1}^{N} \left(H_{E_j} + H_{B_j} +  K_{j,j}^{(1)}\right)\nonumber\\
=& H_{E} + H_B + \sum_{j=1}^{N} i g (\ketbra{1}{0}_{E_j}\otimes \relax{\relax{S}}_{B_j}^{-} - \ketbra{0}{1}_{E_j} \otimes \relax{\relax{S}}_{B_j}^{+}).\label{H_ind}
\end{align}

\subsection{Effective master equation}
The total system of the individual-charging protocol is just $N$ independent and identical copies of a single pair of an engine and a battery, which is included in the collective-charging protocol with $N=1$.
Thus, all we have to analyze is the performance of the collective-charging protocol with generic $N$.
Hence, we focus on the collective-charging protocol.

If the time scale of the thermalization of the engine with each heat bath is much shorter than the time scale of the interaction with the battery, the engine keeps the population inversion and effectively works as a reservoir with negative temperature \cite{PhysRevA.93.052119}.
We can estimate the time scale of the transitions induced by the coupling between the engine and the battery as $\left(\frac{g}{\sqrt{N}} \sqrt{\langle \relax{S}_{B}^{-} \relax{S}_{B}^{+} \rangle + 1}\right)^{-1}$, and the time scale associated with the coupling with each heat bath as $\Gamma_{h(c)}^{-1}[1 + \exp (-\beta_{h(c)} \omega_{h(c)})]^{-1}$.
Since $\langle \relax{S}_{B}^{-} \relax{S}_{B}^{+} \rangle$ is at most of order $O(N^2)$, we have a criterion
\begin{align}
 g \sqrt{N} \ll \Gamma_{h(c)}[1 + \exp (-\beta_{h(c)} \omega_{h(c)})] \label{TS_sep}
\end{align}
for the validity of the ad-hoc master equation (ME) of the state $\rho_B(t)$ of the battery
\begin{align}
 &\frac{d\relax{\rho}_{B}}{dt} \nonumber\\
=& - i[H_B, \rho_B]
 + \Gamma_{e} \left[\mathcal{D}[\relax{\relax{S}}_{B}^{-}](\relax{\rho}_{B}) + e^{-\beta_{e} \omega_{0}} \mathcal{D}[\relax{\relax{S}}_{B}^{+}](\relax{\rho}_{B})\right],\label{eff_ME}
\end{align}
where $\beta_e^{-1} = \omega_0 (\beta_h \omega_h - \beta_c \omega_c)^{-1} < 0$ is the effective temperature, and $\Gamma_e = (2g)^2 (\Gamma_h e^{-\beta_h \omega_h} + \Gamma_c e^{-\beta_c \omega_c})^{-1} (1 + e^{-\beta_h \omega_h} + e^{-\beta_e \omega_0})^{-1}$ is the effective damping rate.
Equation~\eqref{eff_ME} can be derived by heuristically applying the Born-Markov-type approximation based on the time-scale separation \eqref{TS_sep} in a similar way to Appendix B of \cite{PhysRevA.93.052119}.
Especially, the same dynamics $(\Gamma_{e}/N) \left[\mathcal{D}[\relax{\relax{S}}_{B}^{-}](\relax{\rho}_{B}) + e^{-\beta_{e} \omega_{0}} \mathcal{D}[\relax{\relax{S}}_{B}^{+}](\relax{\rho}_{B})\right]$ is induced by each of $N$ engines, and hence ME \eqref{eff_ME} holds in total.

We note that the form of ME \eqref{eff_ME} coincides with the dissipation due to the environment, replacing $\Gamma_e$ and $\beta_e$ with the damping rate $\Gamma$ and the inverse temperature $\beta$ of the environment, respectively.
Hence, in the presence of the environmental dissipation, just the parameters $\Gamma_e$ and $\beta_e$ are modified while keeping the form of ME \eqref{eff_ME}.
The effective damping rate and the inverse temperature are modified to be $\tilde{\Gamma}_e = \Gamma_e + \Gamma$ and $\tilde{\beta}_e = \omega_0^{-1}[\log (\Gamma_e + \Gamma) - \log (\Gamma_e e^{-\beta_e \omega_0} + \Gamma e^{-\beta \omega})]$, respectively.
Thus, as long as $(\Gamma_e e^{-\beta_e \omega_0} + \Gamma e^{-\beta \omega})/(\Gamma_e + \Gamma) > 1$, we still have negative effective temperature $\tilde{\beta_e} < 0$, which means that the charging is stable against the environmental dissipation.
In the following, we investigate the charging performance of the battery initially in the totally empty state $\ket{\psi_0}$ based on ME \eqref{eff_ME}, assuming $\tilde{\beta}_e < 0$, where $\tilde{\Gamma}_e$ and $\tilde{\beta}_e$ will be denoted respectively as $\Gamma_e$ and $\beta_e$ for simplicity of the notation.

\section{Asymptotically perfect charging: Capacity and quality enhancements}\label{sec_capacity}
\subsection{Average charged energy and fluctuation}
\begin{figure*}
\centering
 \includegraphics[clip ,width=7.2in]{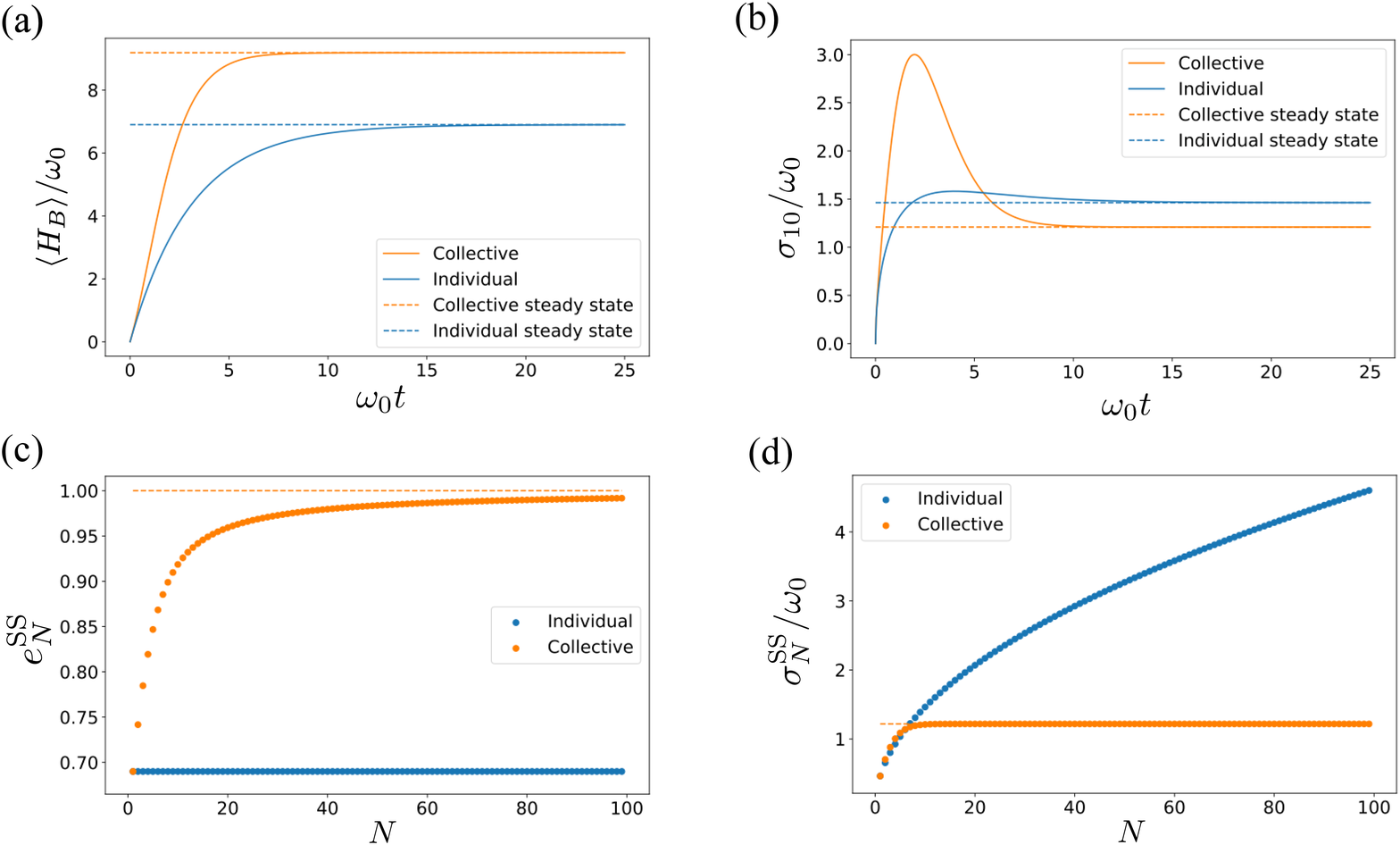}
 \caption{(a) Time evolution of the energy expectation value of the $N = 10$ batteries. (b) Time evolution of the energy fluctuation of the $N=10$ batteries. (c) Energy density of the steady state versus the number $N$ of the batteries. (d) Energy fluctuation of the steady state versus the number $N$ of the batteries. In (a) and (b), orange (blue) curves represent the values of collective (individual) protocol, and orange (blue) dashed lines represent the values in the steady state of the collective (individual) protocol. In (c) and (d), orange (blue) dots represent the values of the collective (individual) protocol. All calculations have been done with the parameters $\beta_e \omega_0 = - 0.8$ and $\Gamma_e / \omega_0 = 0.10$.}
\label{fig-sim}
\end{figure*}
At first, let us look at the numerical results.
We use the Python toolbox ``QuTiP'' \cite{JOHANSSON20131234} to solve ME \eqref{eff_ME}.
Figures~\ref{fig-sim}(a) and \ref{fig-sim}(b) display the time evolution of the expectation value $\langle H_B\rangle := \tr H_B \rho_B$ and the fluctuation \com{$\sigma_{N} := \sqrt{\langle H_B^2\rangle - \langle H_B\rangle^2}$} of the energy respectively, with $N = 10$ for both of the collective and the individual-charging protocols.
The parameters are set to $\beta_e \omega_0 = -0.8$ and $\Gamma_e / \omega_0 = 0.10$.
As can be seen from these figures, the charging ends in convergence to the steady state.
In closed QBs, the charged energy fluctuates in general depending on when we stop the charging \cite{1912.07247}.
Our protocol is intrinsically free from such instability owing to the convergence to the steady state, as in other open QBs \cite{PhysRevApplied.14.024092,2005.08489}.

Figure~\ref{fig-sim}(c) displays the energy density $e_N^{\mathrm{SS}} := \tr H_B \rho_{B,N}^{\mathrm{SS}} / (\omega_0 N)$ of the steady state \com{$\rho_{B,N}^{\mathrm{SS}}$} of $N$ batteries, which stands for the capacity of the battery.
Since each single battery is charged independently in the individual charging protocol, it is obvious that the energy density in this case does not change with the total number $N$ of the batteries.
On the other hand, the energy density in the collective-charging protocol is approaching to the unity, which is the maximum possible value corresponding to the perfect excitation.
Hence, the capacity of the collective $N$ batteries not only scales super-extensively but also approaches to the perfect one in terms of the energy density as $N$ gets large.
Figure~\ref{fig-sim} (d) displays the energy fluctuation $\sigma_{N}^{\mathrm{SS}}$ of the steady state of $N$ batteries, which measures the quality of the charged energy.
It is remarkable that $\sigma_{N}^{\mathrm{SS}}$ immediately converges to a constant around $N=10$ in the collective protocol, in contrast to
the individual protocol, where the fluctuation scales as $\propto \sqrt{N}$.
In fact, these collective enhancements result from emergent quantum statistics induced by the symmetry of ME \eqref{eff_ME}, and hence are universal independently of the parameters.
Below we rigorously show the detail of the collective enhancements.

In the individual protocol, we simply have $N$ independent copies of a battery whose time evolution follows ME \eqref{eff_ME} with $N=1$.
Then, the state of each single battery will converge to the steady state $\rho_{B,1}^{\mathrm{SS}} = (\ketbra{0}_1 + e^{-\beta_e \omega_0} \ketbra{1}_1) / (1 + e^{-\beta_e \omega_0})$.
Hence, the total state becomes $(\rho_{B,1}^{\mathrm{SS}})^{\otimes N}$, and the expectation value and the variance of the energy are just given as $N$ times of $e_1^{\mathrm{SS}}$ and $(\sigma_{1}^{\mathrm{SS}})^2$ respectively.
To investigate the collective protocol, we define collective spin operators
$\relax{S}^{Z}:= \sum_{i=1}^N (\relax{\relax{S}}^{+}_{B_i}\relax{\relax{S}}^{-}_{B_i} - 1/2)$ and $\relax{S}^2 := \frac{1}{2}(\relax{\relax{S}}^{+}_{B}\relax{\relax{S}}^{-}_{B} + \relax{\relax{S}}^{-}_{B}\relax{\relax{S}}^{+}_{B}) + (\relax{S}^{Z})^2$,
though they do not have to represent physical spin degrees of freedom.
We denote the simultaneous eigenstate of $S^{2}$ and $S^{Z}$ with respective eigenvalues $J(J+1)$ and $M$ by $\ket{J, M}$.
From the commutation relations $[S^2, \relax{S}_B^{\pm}] = 0$ and $[S^2, H_B] = 0$, the quantum number $J$ is conserved by the dynamics \eqref{eff_ME}.
Therefore, because the initial state $\ket{\psi_0}$ is $\ket{J = N/2, M = - N/2}$, the time evolution of the battery state $\rho_B(t)$ is contained in the $J=N/2$ subspace $\mathcal{H}_{N/2}$ spanned by $\{\ket{N/2, m - N/2}| m = 0, 1, \cdots, N \}$, a completely symmetric subspace with respect to the permutation of the batteries.
For the analysis of the dynamics within the subspace $\mathcal{H}_{N/2}$, it is convenient to apply the following linear mapping to the $(N+1)$-dimensional Hilbert space spanned by $\{\ket{m}|m = 0, 1, \cdots, N\}$
using the ladder operator $c := \sum_{m=1}^{N}\sqrt{m}\ketbra{m-1}{m}$:
\begin{align}
\pi:&\ket{\frac{N}{2}, m - \frac{N}{2}} \mapsto \ket{m}\\
&\relax{\relax{S}}_B^{-} \mapsto \relax{c}\sqrt{N + 1 - \relax{c}^{\dagger}\relax{c}}\label{pic}\\
&\relax{\relax{S}}_B^{+} \mapsto \left(\sqrt{N + 1 - \relax{c}^{\dagger}\relax{c}}\right)\relax{c}^{\dagger},\label{picd}
\end{align}
which is a finite-dimensional analogy of the Holstein-Primakoff transformation \cite{PhysRev.58.1098}.
In fact, it satisfies $\pi(\relax{S}_{B}^{\pm} \ket{N/2, m - N/2}) = \pi(\relax{S}_{B}^{\pm}) \pi (\ket{N/2, m - N/2})$, and hence we can identify the relevant operators and state vectors through this mapping.
Thus, ME \eqref{eff_ME} with the initial state $\ket{\psi_0}$ is mapped to
\begin{align}
 &\frac{d }{ d t} \pi\left(\relax{\rho}_{B}\right) \nonumber\\
=& - i [\omega_0 c^{\dagger} c, \pi\left(\relax{\rho}_{B}\right)] + \Gamma_e \mathcal{D}\left[\relax{c}\sqrt{N + 1 - \relax{c}^{\dagger}\relax{c}}\right](\pi(\relax{\rho}_{B}))\nonumber\\ 
&+ \Gamma_e e^{-\beta_e \omega_0} \mathcal{D}\left[\left(\sqrt{N + 1 - \relax{c}^{\dagger}\relax{c}}\right)\relax{c}^{\dagger}\right](\pi(\relax{\rho}_{B})).\label{HP_eff}
\end{align}
Following this dynamics, the state converges to the steady state
$\pi(\relax{\rho}_{B,N}^{\mathrm{SS}}) = e^{-\beta_e\omega_0 \relax{c}^{\dagger}\relax{c}}/\tr e^{-\beta_e\omega_0 \relax{c}^{\dagger}\relax{c}}$,
which corresponds to
\begin{align}
 \relax{\rho}_{B,N}^{\mathrm{SS}}
= \frac{1}{Z_N}\sum_{m = 0}^{N} e^{-\beta_e\omega_0 m}\ketbra{\frac{N}{2}, m - \frac{N}{2}},
\end{align}
where $Z_N = \sum_{m = 0}^{N} e^{-\beta_e\omega_0 m}$.
Hence, the steady state turns out to be a mixture of the completely symmetric states $\ket{N/2, m - N/2}$, which is effectively bosonic.
Such an emergent bosonic quantum statistics comes from the $J$ conserving symmetry of the dynamics \eqref{eff_ME} which cannot distinguish each battery.
The significant enhancement in the capacity and the quality of the batteries may be understood as a kind of the Bose-Einstein condensation in this emergent bosonic state where the roles of the ground state and the excited state are now opposite because of the negative temperature.

Indeed, we have analytical expressions of the capacity $N e_{N}^{\mathrm{SS}}$ and the fluctuation $\sigma_N^{\mathrm{SS}}$ of the collective protocol as follows:
\begin{align}
 &N e_N^{\mathrm{SS}} \nonumber\\
=& \tr H_B \relax{\rho}_{B,N}^{\mathrm{SS}} \nonumber\\
=& \frac{N \omega_0}{1 - e^{\beta_e \omega_0 (N + 1)}} - \omega_0 (e^{-\beta_e\omega_0} - 1)^{-1} \frac{1- e^{\beta_e\omega_0 N}}{1- e^{\beta_e\omega_0 (N+1)}},\\
&\sigma_N^{\mathrm{SS}} \nonumber\\
=& \sqrt{\tr H_B^2 \relax{\rho}_{B,N}^{\mathrm{SS}} - (N e_N^{\mathrm{SS}})^2}\nonumber\\
=& \frac{\omega_0 \csch \left(-\frac{\beta_e \omega_0}{2}\right)}{2(1 - e^{\beta_e \omega_0 (N+1) })} + O(N e^{\frac{\beta_e \omega_0}{2} N}).
\end{align}
Since $\beta_e < 0$ and $\omega_0 > 0$, we have $\lim_{N\rightarrow \infty} e_N^{\mathrm{SS}} = 1$ and $\lim_{N\rightarrow\infty} \sigma_N^{\mathrm{SS}} = (\omega_0 /2 ) \csch \left(-\beta_e \omega_0 / 2\right) $.
Therefore, the collective capacity asymptotically approaches the perfect excitation, and
the numerical results are now rigorously proved.
We also note that the speed of the convergence is estimated as $O(N^{-1})$ for $e_N^{\mathrm{SS}}$ and $O(N e^{\frac{\beta_e \omega_0}{2} N})$ for $\sigma_N^{\mathrm{SS}}$, which agrees with Fig.~\ref{fig-sim}(c) and \ref{fig-sim}(d).

\subsection{Ergotropy}
\begin{figure}[!t]
\centering
\includegraphics[clip ,width=3.2in]{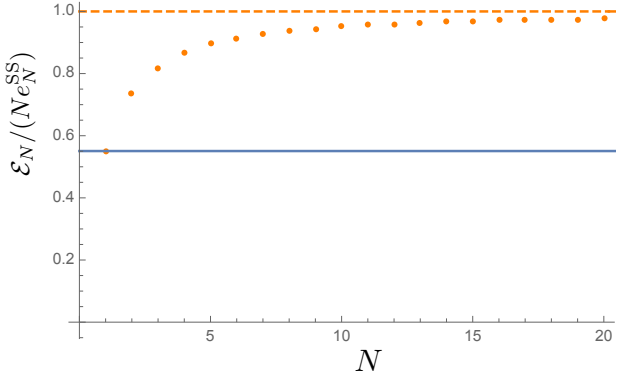}
\caption{Ratio of the ergotropy to the total average charged energy vs $N$. The orange dots are the plot of $\mathcal{E}_N / (N e_N^{\mathrm{SS}})$ for the collective protocol. The solid blue line indicates the ratio $1 - e^{\beta_e \omega_0}$ for the individual protocol. $\beta_e \omega_0 = - 0.8$ is chosen.}
\label{fig_speed}
\end{figure}
In general, the average whole charged energy $N e_N^{\mathrm{SS}}$ is not necessarily available depending on the possible operation.
One reasonable measure of the available energy is the ergotorpy $\mathcal{E}_N:= Ne_N^{\mathrm{SS}} - \min_{U:\mathrm{unitary}}\tr H_B U \rho_{B,N}^{\mathrm{SS}}U^{\dagger}$ \cite{Allahverdyan:2004aa}, which characterizes the maximum extractable energy under a cyclic process where the energy difference is described by a unitary operator.
The optimum unitary operator to extract the energy from $\rho_{B,N}^{\mathrm{SS}}$ is the one which maps $\ket{N/2, N/2}$ to $\ket{\psi_0}$ and $\ket{N/2, m - N/2}$ $(m\neq N)$ to $\ket{0}_1 \otimes \cdots \otimes \ket{0}_{m - 1}\otimes\ket{1}_m\otimes\ket{0}_{m+1} \otimes \cdots \otimes\ket{0}_N$, and hence it turns out that the minimum locked energy asymptotically converges to a constant as follows:
\begin{align}
 &\min_{U:\mathrm{unitary}}\tr H_B U \rho_{B,N}^{\mathrm{SS}}U^{\dagger} \nonumber\\
=& \omega_0 \left(1 - \frac{e^{-\beta_e\omega_0 N}}{Z_N}\right) \nonumber\\
=& \omega_0\left[1 - \frac{1 - e^{\beta_e\omega_0}}{1 - e^{\beta_e\omega_0(N+1)}}\right] \rightarrow \omega_0 e^{\beta_e \omega_0} \quad (N\rightarrow\infty).
\end{align}
Thus, in the collective protocol, the asymptotic freedom ``$\lim_{N\rightarrow\infty} \mathcal{E}_N / (N e_N^{\mathrm{SS}}) = 1$'' \cite{PhysRevLett.122.047702} holds with $O(N^{-1})$ of the speed of convergence.
The asymptotic freedom of the collective protocol implies that almost whole average stored energy $N e_N^{\mathrm{SS}}$ can be extracted if $N$ is sufficiently large.
This is in contrast to the individual protocol,
where the asymptotic freedom is not satisfied because the ergotropy is just extensive as $N \mathcal{E}_1 = - N \omega_0 \tanh (\beta_e \omega_0 /2)$ and hence its ratio to the total energy remains to be a constant value $\mathcal{E}_1 / e_1^{\mathrm{SS}} = 1 - e^{\beta_e \omega_0} < 1$.

In summary, the capacity per single battery in the collective-charging protocol asymptotically reaches the perfect excitation.
That is in contrast to the individual protocol, where the capacity is just extensive.
Moreover, in the collective protocol, the whole average charged energy is actually extractable in the asymptotic sense that the ratio of the ergotropy to the average energy is convergent to unity as $N$ gets large.
That is again in contrast to the individual protocol, where nonzero constant portion of the average energy remains locked regardless of $N$.
These significant enhancements can be understood as a kind of Bose-Einstein condensation to the excited state due to the effective negative temperature and the emergent bosonic quantum statistics of the steady state.
Notably, we have rigorously proved these results based on the analytic solvability of the steady state of our model.


\section{Charging speed and power enhancement}\label{sec_speed}
\begin{figure}[!t]
\centering
\includegraphics[clip ,width=3.2in]{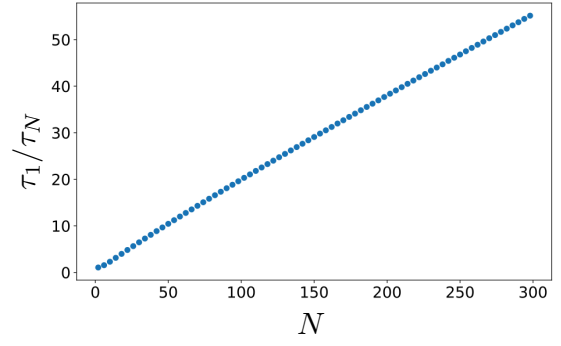}
\caption{\com{Speed rate $\tau_1/\tau_N$ showing an enhancement of the charging speed. The parameters are set to $\beta_e \omega_0 = - 0.8$ and $\Gamma_e/\omega_0 = 0.10$.}}
\label{fig_speed}
\end{figure}
From Figs.~\ref{fig-sim}(a) and \ref{fig-sim}(b), one may guess that the charging speed is also enhanced in the collective protocol.
 This is actually the case as shown in Fig.~\ref{fig_speed}.
We define the charging time
\begin{align}
 \tau_N := \min \left\{t \left| \frac{N e_N^{\mathrm{SS}} - \tr H_B\rho_B(t)}{N e_N^{\mathrm{SS}}} < \epsilon \right.\right\}
\end{align}
at which the batteries are fulfilled up to a small deficit rate $\epsilon$, which is set to $10^{-3}$ in Fig.~\ref{fig_speed}.
Since the individual protocol independently proceeds every battery in the same way, the charging time in the individual protocol is just $\tau_1$ regardless of the total number $N$.
Hence, the ratio $\tau_1 / \tau_N$ quantifies the advantage of the collective protocol in the charging speed.
Notably, in contrast to conventional models, large $N$ is still numerically tractable because the dimension of the Hilbert space to treat is reduced from $2^N$ to $N + 1$ thanks to the correspondence \eqref{HP_eff}.
Figure~\ref{fig_speed} displays almost $\propto N$ behavior of the advantage $\tau_1 / \tau_N$.
\com{Indeed, as shown below, its order of magnitude is at least estimated as $N / \log N$, which is actually almost $N$-order, as it is larger than $N^{1 - \delta}$ order no matter how small $\delta > 0$ is.}

\com{We remark that the collective advantage in the charging power has hence the same scaling of $\sim N$.}
Although $N$-order advantage in the {\it maximum instantaneous} power has been numerically reported in \cite{PhysRevApplied.14.024092} for small numbers $\sim 10$ of batteries, $N$-order advantage in the {\it average} charging power \com{has not been achieved} with only local two-body interactions so far.
In fact, as pointed out by Quach and Munro \cite{PhysRevApplied.14.024092}, $\sqrt{N}$-order collective advantage in the power \cite{PhysRevLett.120.117702,PhysRevA.97.022106,1812.10139,1912.07234} has not been exceeded by local interactions except for the maximum instantaneous power of their model \cite{PhysRevApplied.14.024092}.
Especially, we emphasize that we take the average ``full-charge'' power \com{which is evaluated with respect to the duration $\tau_N$ with which the charging is sufficiently completed.} This is a most practical figure of merit for the charging power.
Furthermore, in the following, we provide a theoretical evidence of this collective enhancement, showing the physics behind it.

\begin{figure}[!t]
\centering
\includegraphics[clip ,width=3.2in]{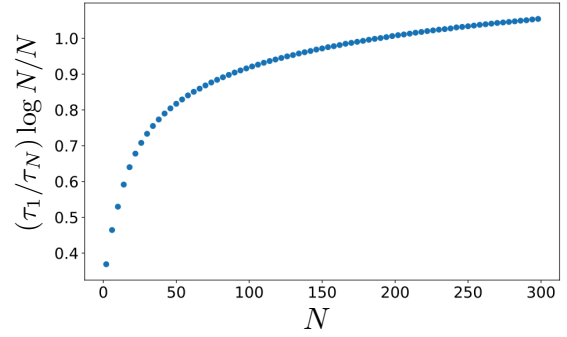}
\caption{\com{Speed rate $\tau_1/\tau_N$ normalized by $N / \log N$ to verify the order of magnitude. The parameters are set to $\beta_e \omega_0 = - 0.8$ and $\Gamma_e/\omega_0 = 0.10$.}}
\label{fig_speed_log}
\end{figure}

The dynamics (\ref{eff_ME}) can be seen as a negative temperature version of the spontaneous emission with the superradiance \cite{PhysRevA.2.2038}.
Actually, the charging speed enhancement can be understood as a similar phenomenon to the superradiance except that the radiation is replaced by the absorption due to the negative temperature.
Since $H_B = \omega_0 S^{Z} + \omega_0 N / 2$, we focus on the time evolution of the expectation value $\langle S^{Z} \rangle (t) = \tr S^{Z}\rho_B(t)$.
ME (\ref{eff_ME}) yields the equation of the time evolution
\begin{align}
 \frac{d}{d t} \langle S^{Z}\rangle = \Gamma_e (e^{-\beta_e \omega_0} - 1) \langle S_B^{-} S_B^{+}\rangle -2\Gamma_e \langle S^{Z}\rangle \label{SZ_eq}
\end{align}
of $S^{Z}$.
As seen from this equation, the correlation $\langle S_B^{-} S_B^{+}\rangle$ among the batteries determines the excitation rate as similar to the superradiance.
Especially, for the state $\ket{N/2, m - N/2}$, the right hand side of \eqref{SZ_eq} reads
\begin{align}
&W_{N,m} \nonumber\\
:=&\Gamma_e (e^{-\beta_e\omega_0} - 1)\bra{N/2, m - N/2}S_B^{-} S_B^{+}\ket{N/2, m - N/2} \nonumber\\
&- 2\Gamma_e \bra{N/2, m - N/2}S^{Z}\ket{N/2, m - N/2}\nonumber\\
 =&\Gamma_e (e^{-\beta_e\omega_0} - 1)\left[(N - m) m +
\frac{N - (1+ e^{\beta_e\omega_0})m}{1-e^{\beta_e\omega_0}}\right].
\end{align}
At $m = N/2$, the excitation rate reaches $N^2$-order in accordance with the superradiance.
Moreover, ME \eqref{eff_ME} guarantees that 
in the whole time evolution from the initial state $\ket{N/2, -N/2}$, the state is kept in some mixture of eigenstates $\ket{N/2, M}$ $(M = -N/2, \cdots, N/2)$ without any offdiagonal term between them.
That is because any eigenstate $\ket{N/2,M}$ evolves to a mixture of eigenstates as verified by substitution.
Thus, $W_{N,m}^{-1}$ gives the time scale of the process $\ket{N/2, m - N/2} \rightarrow \ket{N/2, m - N/2 + 1}$ as long as $W_{N,m} > 0$.
In fact, the portion of $m$ with $W_{N,m} \leq 0$ asymptotically gets negligible as $(N - m_{+}(N))/N \xrightarrow{N\rightarrow \infty} 0$, where $m_{+}(N) := \max\{m|W_{N,m}>0\}$.
Therefore, the total duration of the charging can be roughly estimated by summing up each duration $W_{N,m}^{-1}$ of a single step in the whole cascade of excitations \cite{GROSS1982301}.
Then, the order of magnitude of the duration is estimated as
\begin{align}
 \tau_N \sim& \frac{1}{\Gamma_e (e^{-\beta_e\omega_0}-1)}\sum_{m = 1}^{N - 1} \frac{1}{(N-m)m}\nonumber\\
=&\frac{2}{\Gamma_e (e^{-\beta_e\omega_0}-1)N} \left[1 + \frac{1}{2} + \cdots + \frac{1}{N - 1} \right] \nonumber\\
\sim& \frac{2\log N}{\Gamma_e (e^{-\beta_e\omega_0}-1) N}.
\end{align}
\com{Therefore, the scaling of $\tau_1 / \tau_N$ is roughly estimated as $N/\log N$.
This is actually consistent with the numerical result in Fig.~\ref{fig_speed_log}.}


\section{Conclusion}\label{conclusion}
 We have proposed a model of quantum batteries charged by quantum heat engines (QHEs) and assessed its performance.
Accordingly, our model shows the simultaneous achievability of the asymptotically perfect charge and close to $N$-order charging power enhancement of quantum batteries (QBs) with only thermal energy resource and simple local interactions in a stable manner.
One of the advantages of our model is the simplicity of the mechanisms of the collective enhancements.
The capacity enhancement is caused by a kind of Bose-Einstein condensation to the excited state
due to the effective indistinguishability \com{and the effective negative temperature}.
The collective enhancement of the charging power is explained as a negative temperature version of the superradiance phenomena.
In this sense, these enhancements in our model are kinds of quantum collective effects.
These simple mechanisms behind the collective enhancements may hint for understanding how to obtain high-performance QBs.

An important future work is the experimental implementation of the QBs.
The experimental realization of our model is also meaningful as an implementation of the QHE with its distinct application as 
a high-performance charger of QBs.
\com{Our results suggest that QHEs actually fit for a charger of QBs,
exploiting the collective enhancements, not only converting the disordered thermal energy to the ordered energy stored in quantum degrees of freedom.}
\com{A possible implementation is to use the capacitively-shunted (C-shunted) flux qubit \cite{PhysRevB.75.140515,Yan:2016aa} as the heat engine, and nitrogen–vacancy color centers (NV$^-$ centers) as the QBs.
Because the second-excited level of the C-shunted flux qubit can be accessible \cite{PhysRevB.75.140515}, it is possible to use it as a three-level system.
The coupling with two different heat baths may be realized by the model discussed in \cite{PhysRevB.76.174523}. 
In that model, a flux qubit is coupled and decoupled to two bath-registers each of which is embedded in an LCR circuit acting as a band-pass filter.
In our case, the frequencies of the band-pass filters should be adjusted to the two different modes of the C-shunted flux qubit as the three-level engine.
Moreover, the coherent coupling of a superconducting flux qubit to many NV$^-$ centers has been realized \cite{Zhu:2011aa}.
Hence, all the components of our model are feasible to realize.}


Another future work is further thermodynamic analysis such as heat and efficiency evaluations related to the second law in quantum thermodynamics.
The second law is not trivial in our case since the quantum batteries also have positive entropy production.
That is related to the nontriviality of the definition of the work and heat in quantum heat engines \cite{niedenzu_concepts_2019}.


\begin{acknowledgments}
 We would like to thank Yuichiro Matsuzaki for helpful comments.
\com{This work is supported by the Zhejiang Provincial Natural Science Foundation Key Project (Grant No. LZ19A050001), by National Natural Science Foundation of China (Grants No. 11975199 and No. 11674283), by the Fundamental Research Funds for the Central Universities (2017QNA3005 and 2018QNA3004), and by the Zhejiang University 100 Plan.}
K. I. is also supported by MEXT Quantum Leap Flagship Program (MEXT Q-LEAP) (Grant No. JPMXS0118067394).
\end{acknowledgments}

\bibliographystyle{apsrev4-1}
\bibliography{papers_aps}

\end{document}